\documentclass[shortnote,letterpaper]{jpsj3}
\usepackage[dvips]{graphicx} 
\usepackage{txfonts}
\title{Charge Excitations in Nd$_{2-x}$Ce$_x$CuO$_4$ Observed with Resonant Inelastic X-ray Scattering: Comparison of Cu K-edge with Cu L$_3$-edge}
\author{Kenji Ishii$^1$\thanks{kenji@spring8.or.jp}, Masahito Kurooka$^2$, Yusuke Shimizu$^2$,\\
Masaki Fujita$^3$, Kazuyoshi Yamada$^3$, Jun'ichiro Mizuki$^2$}
\inst{$^1$Synchrotron Radiation Research Center, National Institutes for Quantum and Radiological Science and Technology, Hyogo 679-5148, Japan \\
$^2$Graduate School of Science and Technology, Kwansei Gakuin University, Hyogo 669-1337, Japan \\
$^3$Institute for Materials Research, Tohoku University, Sendai 980-8577, Japan}
\abst{We report a Cu K-edge resonant inelastic x-ray scattering (RIXS) study of momentum-dependent charge excitations in Nd$_{2-x}$Ce$_x$CuO$_4$ ($x$ = 0.075 and 0.18).
The peak position and width of the excitations coincide excellently with those observed in Cu L$_3$-edge RIXS.
It demonstrates that the same charge excitations are observed at the two edges.}

\begin{document}
\maketitle

Low-energy electron dynamics in high-$T_{\mathrm c}$ cuprates is characterized by the motion of charge and spin.
Among the various experimental technique for studying the electron dynamics, resonant inelastic x-ray scattering (RIXS) has gained a great deal of attention because one can measure electronic excitation spectra with momentum resolution and element selectivity.\cite{Ament2011,Ishii2013}
Cu $K$- and $L_3$-edges are mostly used for the RIXS study of the cuprates.
Energy resolution has been improved significantly in the last two decades and it reaches a few tens of  meV at best at the edges.\cite{Ketenoglu2015,Brookes2018a}
While spin excitation (single spin-flip process) is allowed only at the $L_3$-edge, charge excitation can be observed at both edges.

In the electron-doped cuptate Nd$_{2-x}$Ce$_x$CuO$_4$ (NCCO), momentum-dependent charge excitations, which are located at higher energy than the spin excitations, are indeed observed in the K-\cite{Ishii2005b,Ishii2006,Ishii2014} and $L_3$-edge RIXS spectra,\cite{Ishii2014,Lee2014,Hepting2018} even though different interpretations, intraband particle-hole excitations\cite{Ishii2005b,Ishii2014} and a certain mode associated with a symmetry-breaking state,\cite{Lee2014} were proposed.
Recently, the charge excitations in the $L_3$-edge RIXS were found to depend not only on the the in-plane momentum but also on the out-of-plane one and ascribed to a plasmon mode\cite{Hepting2018} which was proposed theoretically.\cite{Greco2016,Greco2019}
In order to further investigate the character of the charge excitations, for example, contrasting behavior at high temperature within a slight difference of carrier concentration ($\sim$0.02 electron per Cu atom),\cite{Lee2014} it is important to verify whether the same excitation is observed at the two edges because each edge has a suited energy-momentum range for observing the excitations.
In Ref.~\citen{Ishii2014}, the charge excitations at the $K$-edge are compared with the one at the $L_3$-edge.
However the comparison is made only at a few momentum points and the out-of-plane momentum is not considered.
In this short note, we report a Cu $K$-edge RIXS study of NCCO in comparison with the $L_3$-edge and conclude that the momentum-dependent charge excitations in the Cu $K$-edge RIXS spectra are the same as in the $L_3$-edge.

RIXS experiments were performed at BL11XU of SPring-8.
Incident x rays were monochromatized by a Si(111) double-crystal monochromator and a Si(444) channel-cut monochromator, and horizontally scattered x rays were analyzed in energy by a Ge(733) analyzer.
Experimental geometry was the same as the previous works;\cite{Ishii2005b,Ishii2014} $\pi$-polarized incident photons with 8991 eV were irradiated on the $ac$-plane of NCCO, but the energy resolution of 100 meV was improved better than the works.\cite{Ishii2005b,Ishii2006,Ishii2014}
All the spectra were taken at 10 K.
We use Miller index ($H$, $K$, and $L$) of the body-centered-tetragonal crystallographic unit cell for momentum transfer (${\mathbf Q}$).

\begin{figure}
\includegraphics[width=\textwidth]{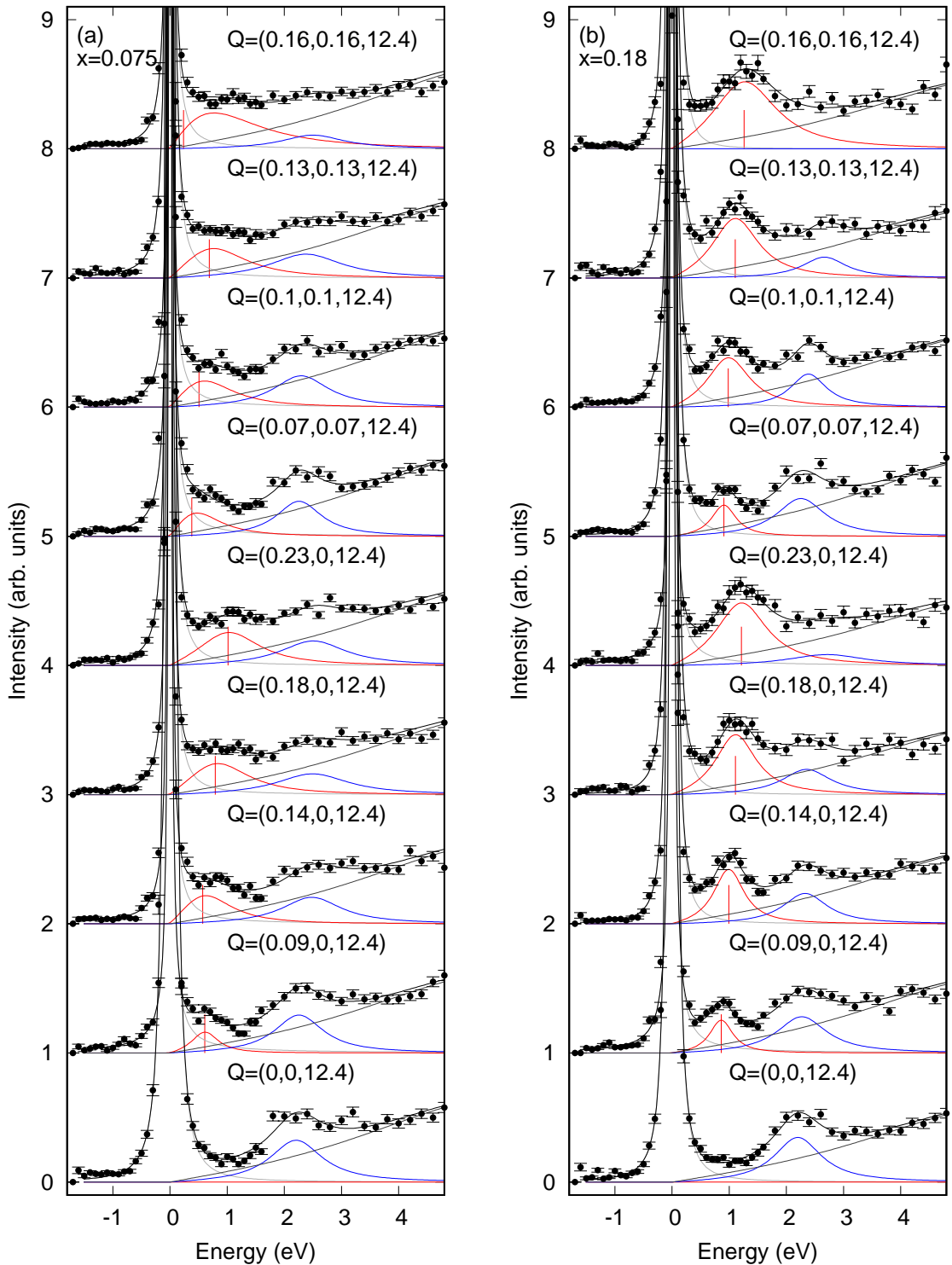}
\caption{(Color Online) RIXS spectra of NCCO for (a) $x$ = 0.075 and (b) $x$ = 0.18.
Filled circles are the experimental data and solid lines are the fitting results of elastic (light gray), momentum-dependent charge excitation (red), interband excitation across the charge-transfer gap (blue), tail of high-energy excitation (dark gray), and sum of the all components (black). Peak position of the momentum-dependent charge excitation is indicated by vertical bars in each spectrum.}
\label{spectrum}
\end{figure}

\begin{figure}
\includegraphics[width=\textwidth]{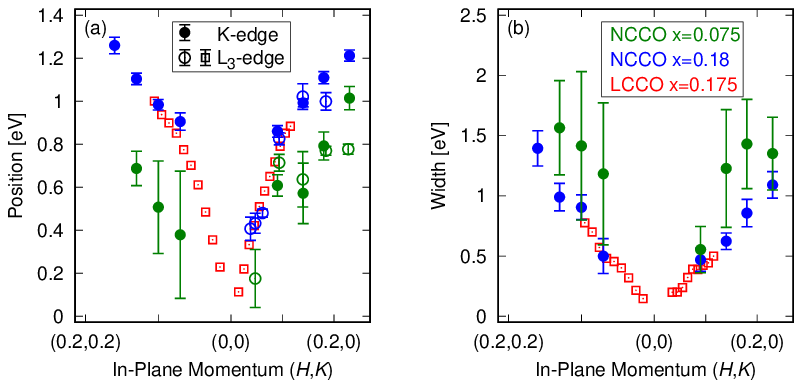}
\caption{(Color Online) Peak position and width of the momentum-dependent charge excitations in the electron-doped cuprates.
Filled circles are the fitting results of the Cu $K$-edge RIXS in Fig.~\ref{spectrum}.
Open circles and squares are those of the Cu $L_3$-edge taken from Refs.~\citen{Ishii2014} and \citen{Hepting2018}, respectively.
The width cannot be evaluated reliably in the spectra of Ref.~\citen{Ishii2014} because the tail of the excitation overlap considerably with the spin and $dd$ excitations due to the low energy resolution and the experimental condition which is suitable for observing the spin excitations.}
\label{peak}
\end{figure}

Figure \ref{spectrum} shows the Cu $K$-edge RIXS spectra of NCCO.
A peak at 2 eV and a momentum-dependent feature below the peak are consistent with the previous work, \cite{Ishii2005b} but the improved energy resolution enables us to observe the latter more clearly, especially at low in-plane momenta ($H$, $K$).
The 2-eV peak is an interband excitation across the charge-transfer gap and it is also observed in parent Nd$_2$CuO$_4$.\cite{Ishii2006}
On the other hand, the momentum-dependent feature appears when electrons are doped.
We fit the spectra by the sum of elastic scattering, the momentum-dependent charge excitation, the interband excitation, and a tail of high energy charge excitation peaked around 6 eV.\cite{Hill1998}
The experimental resolution is considered for the momentum-dependent charge excitation and the interband excitation.
Filled circles in Fig.~\ref{peak}(a) and (b) are the peak position and width (full-width at half maximum) of the momentum-dependent feature obtained from the fitting analysis, respectively.
Because intensity of the feature is roughly proportional to the carrier density, the feature of $x$ = 0.075 forms continuum-like spectral shape in contrast to the salient peak in $x$ = 0.18.
It makes the fitting analysis of $x$ = 0.075 difficult and this is the reason why the error is larger than $x$ = 0.18.

Open circles in Fig.~\ref{peak}(a) are the peak positions in our Cu $L_3$-edge study,\cite{Ishii2014} where 
the crystals from the same batch as this work were used.
Momentum transfer along the ${\mathbf c}^{\ast}$-direction ($L$) varies between 1.55 and 1.65 and it is almost equivalent to that in the present data in Fig.~\ref{spectrum}, considering that the dispersion is folded at even numbers of $L$.\cite{Hepting2018}
The peak positions are consistent between the Cu $K$- and $L_3$-edges.
In addition, we plot the peak positions of La$_{2-x}$Ce$_x$CuO$_4$ (LCCO) of $x$ = 0.175 and $L$ = 1.65 (open squares) taken from a recent high-energy-resolution work at the Cu $L_3$-edge.\cite{Hepting2018}
The positions of LCCO $x$ = 0.175 agree very well with those of NCCO $x$ = 0.18.

In Fig.~\ref{peak}(b), the width of the peak is compared between the $K$-edge and the $L_3$-edge.
The momentum dependence of the width of LCCO $x$ = 0.175 at the $L_3$-edge is connected smoothly to that of NCCO $x$ = 0.18 at the $K$-edge.
It means that the broadening of the peak comes from intrinsic electronic properties of the cuprates rather than some effects in the RIXS process.
While momentum-independent lifetime of electron is phenomenologically taken into account in a theoretical work,\cite{Greco2019} RIXS can provide experimental data to discuss microscopic origin of the broadening quantitatively.

The excellent agreement of the peak position and width between NCCO $x$ = 0.18 and LCCO $x$ = 0.175 proves that the momentum-dependent charge excitation in the Cu $K$-edge RIXS spectra has the same origin as the dispersive mode observed at the Cu $L_3$-edge.
Our result approves the complementary use of the two edges for exploring the charge excitations throughout the energy-momentum space.
In general, the $L_3$-edge is advantageous for the excitations at low energy due to weak elastic scattering while it has limitation of the accessible Brillouin zone.
In the case of cuprates, huge $dd$ excitations above 1.5 eV hampers the observation of the charge excitations at the 	energy.
On the other hand, the $K$-edge does not have the shortcomings of the $L_3$-edge, but it is difficult to measure the inelastic signal at very low energy.
Even though the lower limit in the energy resolution of this study is 0.4-0.5 eV, we will have change to access lower energy if the best resolution (25 meV) is achieved.\cite{Ketenoglu2015}

Another finding is that the peak position of the charge excitations shifts to higher energy with increasing electron doping.
Although the data of $x$ = 0.075 scatter, Fig.~\ref{peak}(a) shows that the peak position of $x$ = 0.18 is higher in energy than that of $x$ =  0.075.
Such doping dependence has been reported in the $L_3$-edge RIXS work\cite{Hepting2018} as a character of plasmon excitation and we confirm it here at the $K$-edge.
It is noted that doping dependence was measured at higher in-plane momentum in the previous $K$-edge work \cite{Ishii2005b,Ishii2006} and it just shows increase of intensity with increasing doping.
It may indicate that the high-energy shift is limited at low in-plane momentum and doping effect of the charge excitations changes at a certain momentum.

In summary, we performed a Cu $K$-edge RIXS study of NCCO and analyzed the momentum-dependent charge excitations below the charge transfer gap.
The peak position and width agree very well with those at the Cu $L_3$-edge and we conclude that the same charge excitations are observed at the two edges.

\begin{acknowledgment}
The authors would like to thank T. Tohyama and H. Yamase for fruitful discussion.
The synchrotron radiation experiments were performed at the BL11XU of SPring-8 with the approval of the Japan Synchrotron Radiation Research Institute (JASRI) (Proposals No.~2013B3502 and 2014A3502).
This work was financially supported by JSPS KAKENHI Grants No.~25400333. and No.~16H04004.
\end{acknowledgment}

\bibliographystyle{jpsj}
\bibliography{ncco-kl}

\begin{thebibliography}{10}

\bibitem{Ament2011}
L.~J.~P. Ament, M.~van Veenendaal, T.~P. Devereaux, J.~P. Hill, and J.~van~den
  Brink, Rev. Mod. Phys. {\bfseries 83},  705 (2011).

\bibitem{Ishii2013}
K.~Ishii, T.~Tohyama, and J.~Mizuki, J. Phys. Soc. Jpn. {\bfseries 82},  021015
  (2013).

\bibitem{Ketenoglu2015}
D.~Ketenoglu, M.~Harder, K.~Klementiev, M.~Upton, M.~Taherkhani, M.~Spiwek,
  F.-U. Dill, H.-C. Wille, and H.~Yava{\c{s}}, J. Synchrotron Radiat.
  {\bfseries 22},  961 (2015).

\bibitem{Brookes2018a}
N.~Brookes, F.~Yakhou-Harris, K.~Kummer, A.~Fondacaro, J.~Cezar, D.~Betto,
  E.~Velez-Fort, A.~Amorese, G.~Ghiringhelli, L.~Braicovich, R.~Barrett,
  G.~Berruyer, F.~Cianciosi, L.~Eybert, P.~Marion, P.~van~der Linden, and
  L.~Zhang, Nucl. Instrum. Methods Phys. Res. A {\bfseries 903},  175  (2018).

\bibitem{Ishii2005b}
K.~Ishii, K.~Tsutsui, Y.~Endoh, T.~Tohyama, S.~Maekawa, M.~Hoesch,
  K.~Kuzushita, M.~Tsubota, T.~Inami, J.~Mizuki, Y.~Murakami, and K.~Yamada,
  Phys. Rev. Lett. {\bfseries 94},  207003 (2005).

\bibitem{Ishii2006}
K.~Ishii, K.~Tsutsui, Y.~Endoh, T.~Tohyama, S.~Maekawa, M.~Hoesch,
  K.~Kuzushita, T.~Inami, M.~Tsubota, K.~Yamada, Y.~Murakami, and J.~Mizuki,
  AIP Conf. Proc. {\bfseries 850},  403 (2006).

\bibitem{Ishii2014}
K.~Ishii, M.~Fujita, T.~Sasaki, M.~Minola, G.~Dellea, C.~Mazzoli, K.~Kummer,
  G.~Ghiringhelli, L.~Braicovich, T.~Tohyama, K.~Tsutsumi, K.~Sato,
  R.~Kajimoto, K.~Ikeuchi, K.~Yamada, M.~Yoshida, M.~Kurooka, and J.~Mizuki,
  Nat. Commun. {\bfseries 5},  3714 (2014).

\bibitem{Lee2014}
W.~S. Lee, J.~J. Lee, E.~A. Nowadnick, S.~Gerber, W.~Tabis, S.~W. Huang, V.~N.
  Strocov, E.~M. Motoyama, G.~Yu, B.~Moritz, H.~Y. Huang, R.~P. Wang, Y.~B.
  Huang, W.~B. Wu, C.~T. Chen, D.~J. Huang, M.~Greven, T.~Schmitt, Z.~X. Shen,
  and T.~P. Devereaux, Nat. Phys. {\bfseries 10},  883 (2014).

\bibitem{Hepting2018}
M.~Hepting, L.~Chaix, E.~W. Huang, R.~Fumagalli, Y.~Y. Peng, B.~Moritz,
  K.~Kummer, N.~B. Brookes, W.~C. Lee, M.~Hashimoto, T.~Sarkar, J.-F. He, C.~R.
  Rotundu, Y.~S. Lee, R.~L. Greene, L.~Braicovich, G.~Ghiringhelli, Z.~X. Shen,
  T.~P. Devereaux, and W.~S. Lee, Nature {\bfseries 563},  374 (2018).

\bibitem{Greco2016}
A.~Greco, H.~Yamase, and M.~Bejas, Phys. Rev. B {\bfseries 94},  075139 (2016).

\bibitem{Greco2019}
A.~Greco, H.~Yamase, and M.~Bejas, Commun. Phys. {\bfseries 2},  3 (2019).

\bibitem{Hill1998}
J.~P. Hill, C.-C. Kao, W.~A.~L. Caliebe, M.~Matsubara, A.~Kotani, J.~L. Peng,
  and R.~L. Greene, Phys. Rev. Lett. {\bfseries 80},  4967 (1998).

\end{thebibliography}
\end{document}